\documentclass[10pt,amsmath,amssymb]{revtex4-1}
\usepackage{textcomp}
\usepackage{graphicx}
\usepackage{longtable}
\usepackage{color}
\usepackage{bm}
\usepackage{dcolumn}

\begin{document}

\title{Anisotropic optical trapping of ultracold erbium atoms}

\author{M. Lepers$^{1}$, J.-F. Wyart$^{1,2}$ and O. Dulieu$^{1}$}
\affiliation{${}^{1}$Laboratoire Aim\'e Cotton, CNRS/Univ.~Paris-Sud/ENS-Cachan, B\^at.~505, Campus d'Orsay, 91405 Orsay, France}
\email{maxence.lepers@u-psud.fr}
\affiliation{${}^{2}$LERMA, UMR8112, Observatoire de Paris-Meudon, Univ.~Pierre et Marie Curie, 92195 Meudon, France}

\date{\today}

\begin{abstract}
Ultracold atoms confined in a dipole trap are submitted to a potential whose depth is proportional to the real part of their dynamic dipole polarizability. The atoms also experience photon scattering whose rate is proportional to the imaginary part of their dynamic dipole polarizability. In this article we calculate the complex dynamic dipole polarizability of ground-state erbium, a rare-earth atom that was recently Bose-condensed. The polarizability is calculated with the sum-over-state formula inherent to second-order perturbation theory. The summation is performed on transition energies and transition dipole moments from ground-state erbium, which are computed using the Racah-Slater least-square fitting procedure provided by the Cowan codes. This allows us to predict 9 unobserved odd-parity energy levels of total angular momentum $J=5$, 6 and 7, in the range 25000-31000 cm$^{-1}$ above the ground state. Regarding the trapping potential, we find that ground-state erbium essentially behaves like a spherically-symmetric atom, in spite of its large electronic angular momentum. We also find a mostly isotropic van der Waals interaction between two ground-state erbium atoms, characterized by a coefficient $C_6^\mathrm{iso}=1760$ a.u.. On the contrary, the photon-scattering rate shows a pronounced anisotropy, since it strongly depends on the polarization of the trapping light.
\end{abstract}

\maketitle

\section{Introduction}

In the field of ultracold atomic and molecular matter, quantum gases composed of particles with a strong intrinsic permanent dipole moment, referred to as dipolar gases, have attracted a lot of interest during the last few years, as they can be manipulated by external electric or magnetic fields \cite{carr2009, dulieu2009, baranov2008, lahaye2009}. Due to the long-range and anisotropic particle-particle interactions, dipolar gases offer the possibility to produce and study highly-correlated quantum matter, which are crucial for quantum information, or for the simulation of many-body or condensed-matter physics \cite{bloch2012, galitski2013}.
The production of ultracold heteronuclear bialkali molecules, which carry a permanent electric dipole moment, in the lowest electronic state \cite{wang2004a, sage2005}, the ground rovibronic \cite{ni2008, deiglmayr2008a} and even hyperfine level \cite{ospelkaus2010b}, was a ground-breaking result, as it demonstrated the possibility to control both the internal and external molecular degrees of freedom \cite{demiranda2011}.

Alternatively open-shell atoms possess a permanent magnetic dipole moment which is determined by their total angular momentum. The latter has the smallest possible value for alkali-metal atoms, namely 1/2, but it can be significantly larger for transition-metal or rare-earth atoms.
In the context of ultracold matter, the first Bose-Einstein condensates of highly-magnetic atoms, obtained with chromium \cite{griesmaier2005, beaufils2008}, were also crucial achievements. Later on, lanthanides started to draw a lot of attention: ultracold erbium atoms were produced in a magneto-optical trap in 2006 \cite{mcclelland2006}. More recently Bose-Einstein condensation was reached with erbium \cite{aikawa2012} and dysprosium \cite{lu2010, youn2010a, youn2010b, lu2011a, lu2011b}, and ultracold thermal samples of thulium \cite{sukachev2010, sukachev2011} and holmium were also produced. These achievements stimulated both theoretical \cite{dzuba2010, dzuba2011, tomza2013, kozlov2013, safronova2013} and experimental studies \cite{newman2011, frisch2013}, which complemented the work on ytterbium, the heavier (closed-shell) lanthanide element (see for example Ch.~1 of \cite{cohen-tannoudji2013} and references therein).

In the present paper we theoretically investigate the optical trapping of ground-state $^3H_6$ erbium atoms. The efficiency of the trapping mechanism relies on the knowledge of the dynamic dipole polarizability, which is a complex quantity depending on the trapping laser frequency $\omega$ and determining the optical potential depth and the photon scattering rate. We compute the dynamic dipole polarizability with a sum-over-state formula, whose versatility enables us to calculate both the real and imaginary parts of the polarizability at any desired frequency.
Two theoretical values of the static ($\omega=0$) dipole polarizability are reported in the literature \cite{chu2007, lide2012}, which were calculated with purely \textit{ab initio} methods. But as shown in recent papers, modeling lanthanides with such methods is a hard task. Here the relevant transition energies from the ground state and the related transition dipole moments are extracted from a semi-empirical approach combining quantum-chemical calculations and experimental data. One central  objective of this article is to determine in which extent the non-spherical electronic distribution of erbium induces an anisotropic response to the trapping light.

Unlike alkali metals, lanthanides are characterized by a complex electronic structure since they possess an open $4f$ and/or $5d$ subshells in their electronic core, which is surrounded by a closed $6s$ shell. Since the electronic angular momentum associated with such configurations is larger, the electronic distribution of a particular Zeeman sublevel is strongly anisotropic. In addition the excitation of the core electrons occurring around 10000 cm$^{-1}$ above the ground-state energy gives birth to very rich and complex spectra whose interpretation was an important part of atomic physics in the last decades \cite{cowan1981, judd1985}. Today the knowledge of the spectroscopy of neutral and charged lanthanides including erbium is still incomplete \cite{wyart2009, denhartog2010, lawler2010}. Therefore using the Racah-Slater least-square fitting method implemented in the Cowan suite of codes \cite{cowan1981}, we adjust calculated and experimental energy levels. This allows us to give a new theoretical interpretation of the spectrum of neutral erbium, and to predict 9 new levels accessible from the ground state through electric-dipole transition.

Since we manipulate a lot of atomic data in this paper, it is necessary to precise how energy levels are labeled. Although an atomic level can be unambiguously defined with its energy with respect to the ground state \cite{ralchenko2009}, information about electronic angular momenta is also crucial. Strictly speaking, the only good quantum numbers are $J$ the total (orbital$+$spin) angular momentum, $M_{J}$ its projection on the quantization axis $z$, and $p$ the parity. For particular states, \emph{e.g.} the lowest states of erbium, the total orbital and spin angular momenta, $L$ and $S$ respectively, are almost good quantum numbers. We also use the leading electronic configuration whose weight depends on the state under consideration (see the Appendix at the end of the paper). For example, ground-state erbium is of even parity and its total angular momentum is $J=6$. It is of $^{3}H$ character ($L=5$, $S=1$) up to 99 \%, the rest being $^{1}I$; its leading configuration is $[\mathrm{Xe}]4f^{12}6s^{2}$. Since our calculations are mostly based on the Wigner-Eckart theorem, we will often label the atomic levels as $|\beta JM_{J}\rangle$, where $\beta$ stands for all quantum numbers except $J$ and $M_{J}$.

The paper is organized as follows. In Section \ref{sec:opt-trap} we give all the formulas necessary to characterize the optical trapping of non-spherically-symmetric atoms, in particular the potential depth and the photon-scattering rate induced by the trapping light. Section \ref{sec:Er-Spec} is dedicated to the spectroscopy of erbium. We first recall the main steps of the present approach based on the Cowan suite of codes, and we present our results for energies and transition dipole moments. In section \ref{sec:pola} we report on  our tests and results for the polarizabilities of ground-state erbium. The reader interested in the final results is invited to go to subsection \ref{sub:pola-res}. Finally section \ref{sec:conclu} contains concuding remarks, emphasizing on the van der Waals interactions between two erbium atoms. More details on the atomic structure calculations are reported in a final Appendix including tables for fitting parameters used to model the erbium spectrum, energies, Land\'e factors, and configuration weights.

\section{Optical trapping of non-spherical atoms \label{sec:opt-trap}}

When spherically-symmetric atoms, like $^{2}S$ alkali-metal or $^{1}S$ alkaline-earth atoms, are submitted to a light wave of angular frequency $\omega$ and intensity $I(\mathbf{r})$, with $\mathbf{r}$ the atomic center-of-mass position in the lab frame $xyz$, $z$ being the quantization axis, they experience a potential energy \cite{grimm2000}
\begin{equation}
U(\mathbf{r};\omega) = -\frac{1}{2\epsilon_{0}c} \Re[\alpha_{\mathrm{scal}}(\omega)]\times I(\mathbf{r})\,,
\label{eq:U}
\end{equation}
which is due to the second-order ac Stark effect. In Eq.~(\ref{eq:U}), $\alpha_{\mathrm{scal}}(\omega)$ is the (complex) scalar dynamic dipole polarizability of the atom, $\Re[...]$ denoting the real part, $\epsilon_{0}$ is the vacuum permitivity and $c$ the speed of light. The presence of the electromagnetic field also induces photon scattering with a rate equal to \cite{grimm2000}
\begin{equation}
\Gamma(\mathbf{r};\omega) = \frac{1}{\hbar\epsilon_{0}c} \Im[\alpha_{\mathrm{scal}}(\omega)]\times I(\mathbf{r})\,,
\label{eq:Gamma}
\end{equation}
where now $\Im[\alpha_{\mathrm{scal}}(\omega)]$ is the imaginary
part of the dynamic scalar dipole polarizability.

The complex polarizability is calculated by using the second-order time-dependent perturbation theory, which is cautiously discussed in Ref.~\cite{langhoff1972}, and by attributing to each excited level a complex energy $E_{\beta'J'}-i\hbar\gamma_{\beta'J'}/2$, $\gamma_{\beta'J'}$ being the inverse lifetime of the level \cite{vexiau2011}. This gives
\begin{equation}
\alpha_{\mathrm{scal}}(\omega) = \frac{1}{3(2J+1)} \sum_{\beta'J'}\left(\frac{\left\langle \beta'J'\right\Vert \mathrm{d}\left\Vert \beta J\right\rangle ^{2}}{E_{\beta'J'}-E_{\beta J}-i\frac{\hbar\gamma_{\beta'J'}}{2}-\hbar\omega}+\frac{\left\langle \beta'J'\right\Vert \mathrm{d}\left\Vert \beta J\right\rangle ^{2}}{E_{\beta'J'}-E_{\beta J}-i\frac{\hbar\gamma_{\beta'J'}}{2}+\hbar\omega}\right)
\end{equation}
with $\langle\beta'J'\Vert\mathrm{d}\Vert\beta J\rangle$ the reduced transition dipole moment. Then considering that the laser frequency is far from any atomic resonance, namely $(E_{\beta'J'} - E_{\beta J} - \hbar\omega) \gg \hbar\gamma_{\beta'J'}/2$, and \textit{a fortiori} $(E_{\beta'J'} - E_{\beta J} + \hbar\omega) \gg \hbar\gamma_{\beta'J'}/2$ since the atoms are in the ground state, we can separate real and imaginary parts
\begin{eqnarray}
\Re[\alpha_{\mathrm{scal}}(\omega)] & = & \frac{2}{3(2J+1)}\sum_{\beta'J'}\frac{\left(E_{\beta'J'}-E_{\beta J}\right)\left\langle \beta'J'\right\Vert \mathrm{d}\left\Vert \beta J\right\rangle ^{2}}{\left(E_{\beta'J'}-E_{\beta J}\right)^{2}-\hbar^{2}\omega^{2}} \label{eq:alpha-scal-re} \\
\Im[\alpha_{\mathrm{scal}}(\omega)] & = & \frac{1}{3(2J+1)} \sum_{\beta'J'} \frac{\left(E_{\beta'J'}-E_{\beta J}\right)^{2}+\hbar^{2}\omega^{2}} {\left[\left(E_{\beta'J'}-E_{\beta J}\right)^{2}-\hbar^{2}\omega^{2}\right]^{2}} \hbar\gamma_{\beta'J'} \left\langle \beta'J'\right\Vert \mathrm{d}\left\Vert \beta J\right\rangle ^{2}.
\label{eq:alpha-scal-im}
\end{eqnarray}
 
For non-spherically-symmetric atoms like erbium, the ac-Stark shift depends on the magnetic sublevel $M_{J}$ and on the light polarization.
In the general case of an elliptically-polarized light whose unit vector of polarization is $\mathbf{e}$, the trapping potential equals \cite{manakov1986}
\begin{eqnarray}
U_{M_{J}}^\mathrm{ell}(\mathbf{r};\theta_{p},\theta_{k},\mathcal{A};\omega) & = & -\frac{1}{2\epsilon_{0}c}I(\mathbf{r})\left\{ \Re[\alpha_{\mathrm{scal}}(\omega)]+\mathcal{A}\cos\theta_{k}\frac{M_{J}}{2J}\Re[\alpha_{\mathrm{vect}}(\omega)]\right.\nonumber \\
 &  & \left.+\frac{3M_{J}^{2}-J(J+1)}{J(2J+1)}\times\frac{3\cos^{2}\theta_{p}-1}{2}\Re[\alpha_{\mathrm{tens}}(\omega)]\right\} ,
\label{eq:U-ell}
\end{eqnarray}
where $\theta_{p}$ is such that $|\mathbf{e}\cdot\mathbf{e}_{z}|^{2}=\cos^{2}\theta_{p}$, $\theta_{k}$ is the angle between $z$ and the wave vector, and $\mathcal{A}$ the ellipticity parameter. Similarly to Eqs.~(\ref{eq:U}) and (\ref{eq:Gamma}) the photon-scattering rate $\Gamma^\mathrm{ell}_{M_{J}}$ is obtained by replacing $\Re[...]$ by $\Im[...]$ in Eq.~(\ref{eq:U-ell}). The quantities $\alpha_{\mathrm{vect}}(\omega)$ and $\alpha_{\mathrm{tens}}(\omega)$ are respectively the vector and  tensor dynamic dipole polarizabilities, given by
\begin{eqnarray}
\Re[\alpha_{\mathrm{vect}}(\omega)] & = & 2\sum_{\beta'J'} X_{JJ'}^{(1)} \frac{\hbar\omega\left\langle \beta'J'\right\Vert \mathrm{d}\left\Vert \beta J\right\rangle ^{2}} {\left(E_{\beta'J'}-E_{\beta J}\right)^{2}-\hbar^{2}\omega^{2}}
\label{eq:alpha-vect-re} \\
\Im[\alpha_{\mathrm{vect}}(\omega)] & = & 2\sum_{\beta'J'}X_{JJ'}^{(1)} \frac{\hbar^2\omega\gamma_{\beta'J'}\left(E_{\beta'J'}-E_{\beta J}\right)} {\left[\left(E_{\beta'J'}-E_{\beta J}\right)^{2}-\hbar^{2}\omega^{2}\right]^{2}} \left\langle \beta'J'\right\Vert \mathrm{d}\left\Vert \beta J\right\rangle ^{2}
\label{eq:alpha-vect-im} \\
\Re[\alpha_{\mathrm{tens}}(\omega)] & = & 4\sum_{\beta'J'}X_{JJ'}^{(2)} \frac{\left(E_{\beta'J'}-E_{\beta J}\right)\left\langle \beta'J'\right\Vert \mathrm{d}\left\Vert \beta J\right\rangle ^{2}} {\left(E_{\beta'J'}-E_{\beta J}\right)^{2}-\hbar^{2}\omega^{2}}
\label{eq:alpha-tens-re} \\
\Im[\alpha_{\mathrm{tens}}(\omega)] & = & 2\sum_{\beta'J'}X_{JJ'}^{(2)} \frac{\left(E_{\beta'J'}-E_{\beta J}\right)^{2}+\hbar^{2}\omega^{2}} {\left[\left(E_{\beta'J'}-E_{\beta J}\right)^{2}-\hbar^{2}\omega^{2}\right]^{2}} \hbar\gamma_{\beta'J'}\left\langle \beta'J'\right\Vert \mathrm{d}\left\Vert \beta J\right\rangle ^{2},
\label{eq:alpha-tens-im}
\end{eqnarray}
where $X_{JJ'}^{(k)}$ are angular factors \cite{angel1968}
\begin{eqnarray}
X_{JJ'}^{(1)} & = & \left(-1\right)^{J+J'} \sqrt{\frac{6J}{(J+1)(2J+1)}}\left\{ 
\begin{array}{ccc}
1 & 1 & 1\\
J & J & J'
\end{array}\right\}
\label{eq:X1} \\
X_{JJ'}^{(2)} & = & \left(-1\right)^{J+J'}\sqrt{\frac{5J(2J-1)} {6(J+1)(2J+1)(2J+3)}} \left\{ \begin{array}{ccc}
1 & 1 & 2\\
J & J & J'\end{array}\right\}.
\label{eq:X2}
\end{eqnarray}

The particular case of a circular right (left) polarization is obtained by setting $\mathcal{A}=+1$ (-1) in Eq.~(\ref{eq:U-ell}). In a linearly-polarized light, corresponding to $\mathcal{A}=0$, the trapping depends neither on the angle $\theta_k$ nor on the vector polarizability. In this case $\theta_p=\theta$ is the angle between the polarization vector $\mathbf{e}$ and the quantization axis $z$. The trapping potential $U_{M_J}^{\mathrm{lin}}$ is obtained from Eq.~(\ref{eq:U-ell})
\begin{equation}
U_{M_{J}}^{\mathrm{lin}} (\mathbf{r};\theta;\omega) = U_{M_{J}}^{\mathrm{ell}} (\mathbf{r};\theta_p=\theta,\theta_k,\mathcal{A}=0;\omega)
\label{eq:U-lin}
\end{equation}
and similarly $\Gamma_{M_{J}}^{\mathrm{lin}} (\mathbf{r};\theta;\omega) = \Gamma_{M_{J}}^{\mathrm{ell}} (\mathbf{r};\theta_p=\theta,\theta_k,\mathcal{A}=0;\omega)$.

\section{Theoretical interpretation of neutral erbium spectrum \label{sec:Er-Spec}}

Equations (\ref{eq:alpha-scal-re}), (\ref{eq:alpha-scal-im}) and (\ref{eq:alpha-vect-re})--(\ref{eq:alpha-tens-im}) above show that the polarizabilities crucially depend on the transition energies and transition dipole moment from erbium ground state. Therefore the quality of those data as well as the method to calculate them represent a central issue of this work.

The initial steps that led to the critical compilation of erbium energy levels were summarized by Martin \textit{et al.}~in \cite{martin1978} and later reported in the NIST database \cite{ralchenko2009}. After 1978, systematic studies of hyperfine effects in $4f^n5d6s6p$ configurations of neutral lanthanides addressed the case of neutral erbium (Er I); but the fine-structure study preceding the determination of magnetic dipole and electric quadrupole parameters had to be limited to the terms of $4f^{11}5d6s6p$ arising from the ground term $^4I$ of the core \cite{kronfeldt1993}.

A first step in the description of Er I levels by means of the Cowan suite of codes \cite{cowan1981} was used in an experimental determination of transition probabilities \cite{lawler2010}. The Cowan codes led to energies and eigenfunctions by least squares determination of radial parameters in appropriate sets of interacting electronic configurations, following the Racah-Slater method as reminded in \cite{wyart2010}. The case of Er I turned out to be more complex than singly-ionized erbium (Er II) \cite{wyart2009}, because in neutral lanthanides the lower levels of many excited configurations overlap the upper part of low-lying configurations. Before 1976, some levels with $7s, 8s, 6d$ electrons were identified by means of very selective decays (from $4f^{12}6s(7s,8s,6d)$ to $4f^{12}6s6p$ and from $4f^{11}5d6s7s$ to $4f^{11}5d6s6p$ and of hazy emission line profiles that are common for such transitions in lanthanides. The semi-empirical designations were tabulated in \cite{martin1978, ralchenko2009}. As concerns combinations of valence electrons, estimates by Brewer \cite{brewer1971, brewer1983} place the lowest levels of odd parity configurations $4f^{12}5d6p, 4f^{11}5d^3, 4f^{11}6s6p^2$, and $4f^{13}6s$ in the energy range 37000-43000 cm$^{-1}$ above the ground level  $4f^{12}6s^2$ $^3H_6$. In the even parity, $4f^{11}5d^26p, 4f^{12}6p^2$ and $4f^{12}5d^2$ should be present above 38500 $\pm 2000$ cm$^{-1}$. The three unknown configurations with $4f^{11}$ core totalize 10914 predicted levels and the four others 1258 levels. The limitations imposed by available computers are less tight than in earlier studies but the applicability of the parametric fitting in the Racah-Slater method is decreased when thousands of adjustable parameters are introduced by several tens of configurations.
This guided us in the choice of the electronic configurations included in the model. In the even parity, since we focus on Er I ground state, we consider the lowest configuration $4f^{12}6s^2$. In the odd parity, we added the high-lying electronic configurations $4f^{12}5d6p$ and $4f^{13}6s$ to the known low-lying ones $4f^{11}5d6s^2$, $4f^{12}6s6p$ and $4f^{11}5d^26s$.

Let us briefly recall the principle of the calculations with the Cowan suite.
\begin{enumerate}
\item First for each configuration separately, the RCN program calculates the electronic wave functions using the relativistic Hartree-Fock (HFR) method.
\item Then RCN2 calculates various radial integrals including: for a given configuration, the direct and exchange Coulombic integrals $F^k(n\ell n'\ell')$ and $G^k(n\ell n'\ell')$ (for equivalent and non-equivalent electrons), the spin-orbit energy $\zeta_{n\ell}$ for each subshell; for each couple of configurations, the configuration-interaction Coulombic integrals $R^k(n\ell n'\ell',n''\ell''n'''\ell''')$. The radial integrals are treated in step (\ref{item:RCE}) as adjustable parameters.
\item \label{item:RCG} Using those radial integrals, RCG diagonalizes the atomic Hamiltonian in appropriate angular-momentum bases, \textit{e.g.} given values of $J$. From the resulting eigenenergies and eigenvectors, it models the atomic spectrum by calculating in particular the Einstein coefficient for all possible electric-dipole transitions.   
\item \label{item:RCE} The energies calculated by RCG are then compared to the tabulated experimental levels. A fit on the atomic parameters is performed by the RCE code in order to minimize the mean error between experimental and theoretical energies. It produces a new set of parameters which serves as input for RCG (step (\ref{item:RCG})). Then a few RCG-RCE loops are performed to minimize the mean error.
\end{enumerate}
We used both the LANL \cite{cowan-lanl} and the Kramida \cite{cowan-kramida} versions of Cowan codes.

The optimal atomic parameters, which make the input for the last call of RCG, are given in appendix (see Tables \ref{tab:parev} and \ref{tab:parev-ci}). In the odd parity, 208 levels of the mixed configurations $4f^{11}5d6s^2$, $4f^{11}5d^26s$, $4f^{12}6s6p$ are used  to determine 24 free parameters, 88 other parameters being constrained. The mean error is 65 cm$^{-1}$. The results are given in the appendix (see Tables \ref{tab:niver1e} and \ref{tab:niver1o}). The general agreement between experimental and theoretical Land\'e factors is a first indication of the quality of the eigenfunctions. The only noticeable exception is the inversion of the close $J=5$ levels at 28026 and 28129 cm$^{-1}$.

\section{Calculation of erbium polarizabilities \label{sec:pola}}

\subsection{Data sets of transition energies and dipole moments}

The output of the previous calculations consists in a list of transition energies and Einstein coefficients, hence of reduced transition dipole moments, from ground-state erbium, which can be used in our sum-over-state formulas of the polarizabilities (see Eqs.~(\ref{eq:alpha-scal-re}), (\ref{eq:alpha-scal-im}), (\ref{eq:alpha-vect-re}--\ref{eq:alpha-tens-im})). We call that list the data set T (after {}``theoretical'').

The data set T has been optimized so that the calculated energies match as well as possible the experimental ones. To obtain better values of the polarizabilities, we apply a second step of optimization by adjusting the monoelectronic radial integrals $\langle n'\ell'|\hat{r}|n\ell\rangle$
to minimize the standard error on Einstein coefficients $A_{i}$
\begin{equation}
\sigma = \left( \sum_{i=1}^{N_{\mathrm{lev}}} \frac{\left(A_{i}^{\mathrm{th}}-A_{i}^{\mathrm{exp}}\right)^{2}} {N_{\mathrm{lev}}-N_{\mathrm{par}}} \right)^{1/2},
\label{eq:std-err-A}
\end{equation}
where $N_{\mathrm{lev}}$ and $N_{\mathrm{par}}$ are the numbers
of levels and adjusted parameters respectively. In this section we discuss in details the influence on (\ref{eq:std-err-A}) of the scaling factors
\begin{equation}
f_{n\ell n'\ell'}=\frac{\langle n'\ell'|\hat{r}|n\ell\rangle}{\langle n'\ell'|\hat{r}|n\ell\rangle_{\mathrm{RHF}}}
\label{eq:SF}
\end{equation}
with $\langle n'\ell'|\hat{r}|n\ell\rangle_{\mathrm{RHF}}$ the relativistic Hartree-Fock radial integral calculated by the Cowan code RCN2.

In order to evaluate the reliability of the data set T, we also consider the 33 lines ending in ground-state erbium which were experimentally detected by Lawler and coworkers \cite{lawler2010}; we obtain the data set E (after {}``experimental''). Theory and experiment can be directly compared by extracting among the lines of data set T those which have an experimental counterpart; this gives the data set T'.

\subsection{Convergence and uncertainty}

Now we discuss the convergence and reliability of our calculations, taking mostly the example of the real part of the static scalar polarizability $\Re[\alpha_{\mathrm{scal}}(\omega=0)]$ (see Eq.~(\ref{eq:alpha-scal-re})). This quantity is not relevant in the context of optical trapping (except for CO$_2$-laser traps) but there exists two theoretical values in the literature to which our results can be compared: 153 a.u.~from Ref.~\cite{lide2012} and 166 a.u.~from Ref.~\cite{chu2007}. The conclusions drawn for $\omega=0$ can actually be extended up to the first main resonances $\omega\lesssim20000$ cm$^{-1}$. Note that the imaginary part of the static scalar polarizability will be examined separately.

\subsubsection{Influence of data sets and scaling factors.}

\begin{table}
\caption{Static scalar dipole polarizability
$\Re[\alpha_{\mathrm{scal}}(\omega=0)]$ for the different data sets and different scaling factors $f$ for the mono-electronic radial integrals (see text).
\label{tab:pola-scal-re-stat}}
\begin{ruledtabular}
\begin{tabular}{ccc}
data set & scaling factor $f$ & $\Re[\alpha_{\mathrm{scal}}(0)]$\tabularnewline
\hline
E  &    - & 132 \tabularnewline
T' &    1 & 200 \tabularnewline
T  &    1 & 226 \tabularnewline
T  & 0.81 & 148 \tabularnewline
T  & 0.77 & 134 \tabularnewline
\end{tabular}
\end{ruledtabular}
\end{table}

First, the influence of the different data sets with no adjustment on the scaling factors $(f_{n\ell n'\ell'}=1)$ is addressed (see first three lines of Table \ref{tab:pola-scal-re-stat}). The two theoretical values T and T' clearly exceed the experimental one E. To have better an agreement between E and T', we should use radial scaling factors smaller than unity. In addition since for zero or weak frequencies, the scalar polarizability {[}Eq.~(\ref{eq:alpha-scal-re}){]} is a sum of positive terms, and since the set of experimental lines is \emph{a priori} incomplete, the value from data set E (132 a.u.) can be regarded as the lower bound for $\alpha_{\mathrm{scal}}(0)$.

Given the configurations of odd parity that we include in our calculation of erbium spectrum (see Sec.~\ref{sec:Er-Spec}), the transition-dipole-moment matrix elements involve two radial integrals: $\langle4f|\hat{r}|5d\rangle$ and $\langle6s|\hat{r}|6p\rangle$. We calculated the standard error on Einstein coefficients (\ref{eq:std-err-A}) between sets E and T', and we found that: (i) it is much less sensitive to $f_{4f,5d}$ than to $f_{6s,6p}$; (ii) the standard error is minimum ($\sigma=1.36\times10^{7}$ s$^{-1}$ with $N_{\mathrm{lev}}=33$ and $N_{\mathrm{par}}=2$) for $f_{6s,6p}=0.77$. The corresponding polarizability is 118 and 134 a.u.~for data sets T' and T respectively. A closer look at the result shows that this scaling factor minimizes the error on the strongest line, whose upper level is the one of $J=7$ at 24943 cm$^{-1}$.

We made another test by searching the factor $f$ giving the same result for the data sets T' and E. We found $f=0.81$ and the corresponding polarizability 148 a.u.. The two {}``optimal'' scaling factors ($f=0.77$ and 0.81) are rather close to each other. Their discrepancy can be explained because the second criterion allows for compensation effects between theoretical Einstein coefficient larger and smaller than the experimental ones.

In conclusion, we take the previous results as our lower and upper bonds, and we take $f=0.79$ for our recommended value. Finally we obtain $\Re[\alpha_{\mathrm{scal}}(\omega=0)]=141\pm7$ a.u., which is in a good agreement although smaller than the two literature values.

\subsubsection{Convergence on excited energy levels.}

Now that the question of scaling factors is solved, we discuss the convergence of the sum-over-state formulas inside the list of excited states in data set T. Namely, we truncate Eqs.~(\ref{eq:alpha-scal-re}), (\ref{eq:alpha-scal-im}), (\ref{eq:alpha-tens-re}) and (\ref{eq:alpha-tens-im}) up to a given excited state $|N\rangle=|\beta_{N}J_{N}\rangle$ of energy $E_{N}$, and we plot the reulting polarizabilities $\alpha_{\mathrm{scal}}^{N}(\omega=0)$ and $\alpha_{\mathrm{tens}}^{N}(\omega=0)$ as functions of $E_{N}$ (note that $\alpha_{\mathrm{vecl}}(\omega=0)=0$).

\begin{figure}
\begin{centering}
\includegraphics[width=8cm]{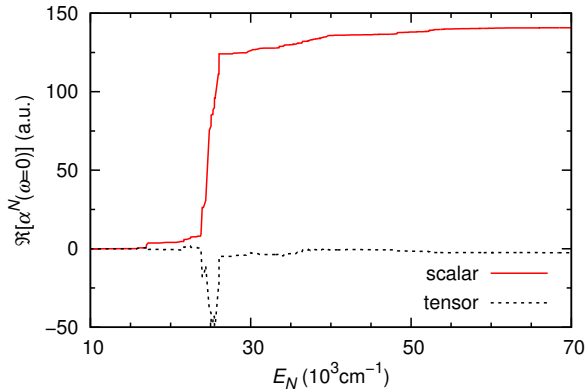}
\par\end{centering}
\caption{(Color online) Convergence of the real part of the static scalar polarizability $\Re[\alpha_{\mathrm{scal}}^{N}(\omega=0)]$ (solid line), and the static tensor polarizability $\Re[\alpha_{\mathrm{tens}}^{N}(\omega=0)]$ (dotted line), with respect to the excited states of data set T. Eqs.~(\ref{eq:alpha-scal-re}) and (\ref{eq:alpha-tens-re}) are truncated up to the excited state of energy $E_{N}$.
\label{fig:pola-real-conv}}
\end{figure}

On Fig.~\ref{fig:pola-real-conv} we focus on the real part of the scalar and tensor polarizabilities. We see that they are converged for $E_{N}\approx60000$ cm$^{-1}$, where they reach at least 99 \% of their total value. In addition, the scalar polarizability reaches already 90 \% of its total value at $E_{N}\approx30000$ cm$^{-1}$, that is when the strongest lines have been included in the sum.
In comparison, the lowest energies associated with configurations not included in our calculation are estimated around 40000 cm$^{-1}$ above the ground state \cite{brewer1971, brewer1983}.
The convergence is visible on the tensor polarizability, although less spectacular, because the angular factor $X_{JJ'}^{(2)}$ {[}Eq.~(\ref{eq:X2}){]} can change sign with $J'$. This fast convergence is due to the $(E_{\beta'J'}-E_{\beta J})^{-1}$ factor in Eq.~(\ref{eq:alpha-vect-re}) which enhances the importance of low-energy transitions; it is also inherent to the erbium spectrum which is composed of a few strong lines among a forest of weak lines.

\subsubsection{Imaginary part of the static scalar polarizability.}

In order to calculate the imaginary part of the polarizabilities, we need, in addition to transition energies and transition dipole moments, the lifetimes of all the excited states. For a given state this would require to know the Einstein coefficient of all the downward transitions from this state. Here we will rather make the assumption that all the excited states can only decay to the ground state. The inverse lifetime of the state $|\beta'J'\rangle$ is then the Einstein coefficient of the transition to the ground state
\begin{equation}
\gamma_{\beta'J'} = \frac{E_{\beta'J'}^{3} \left\langle \beta'J'\right\Vert \mathrm{d}\left\Vert^3H_6\right\rangle ^{2}}{3(2J'+1)\pi\epsilon_0\hbar^4 c^3}
\label{eq:lifetime-approx}
\end{equation}
where we have set the ground-state to zero.
We can check this hypothesis by calculating the imaginary part of the static polarizabilities with the data set E. On the one hand we use Eq.~(\ref{eq:lifetime-approx}) and on the other hand we used the lifetimes measured by the same group \cite{denhartog2010}. We obtain a very good agreement between the two methods which respectively give $\Im[\alpha_\mathrm{scal}(0)]=1.42\cdot 10^{-6}$ and $1.43	\cdot 10^{-6}$ a.u., $\Im[\alpha_\mathrm{tens}(0)]=-3.33\cdot 10^{-7}$ and $-3.41\cdot 10^{-7}$a.u. (note that $\Im[\alpha_\mathrm{vect}(0)]=0$). Indeed Eq.~(\ref{eq:lifetime-approx}) is a very good approximation for the lowest excited states, which are prevailing due to the $E_{\beta'J'}^{-2}$ dependence of Eqs.~(\ref{eq:alpha-scal-im}), (\ref{eq:alpha-vect-im}) and (\ref{eq:alpha-tens-im}).

\begin{figure}
\begin{centering}
\includegraphics[width=8cm]{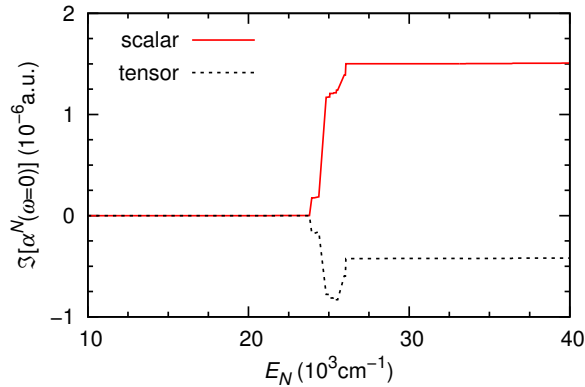}
\par\end{centering}
\caption{(Color online) Convergence of the imaginary part of the static scalar polarizability $\Im[\alpha_{\mathrm{scal}}^{N}(\omega=0)]$ (solid line), and the static tensor polarizability $\Im[\alpha_{\mathrm{tens}}^{N}(\omega=0)]$ (dashed line), with respect to the excited states of data set T. Eqs.~(\ref{eq:alpha-scal-im}) and (\ref{eq:alpha-tens-im}) are truncated up to the excited state of energy $E_{N}$.
\label{fig:pola-imag-conv}}
\end{figure}

Figure \ref{fig:pola-imag-conv} shows the imaginary part of the
static polarizabilities as a function of $E_{N}$.  For the sake of coherence we have used the same scaling factor $f=0.79$ as for the real part of the polarizabilities. The convergence with $E_N$ is even faster than for the real part: at $E_N=30000$ cm$^{-1}$, $\Im[\alpha^N_\mathrm{scal}(0)]$ and $\Im[\alpha^N_\mathrm{tens}(0)]$ differ by less than 1 \% from their final values given in Table \ref{tab:pola-re-im-freq}.

\subsection{Results \label{sub:pola-res}}

In order to present our results in a convenient way for experimental purposes, we give the polarizabilities of erbium in atomic units (units of $a_{0}^{3}$, with $a_{0}$ the Bohr radius), but also the corresponding relevant quantities in physical units. To the real part of the polarizability corresponds the trapping potential in temperature units 
\begin{equation}
U(\mathrm{in\, K}) = \frac{2\pi a_{0}^{3}}{k_{B}c} \Re[\alpha(\mathrm{in\, a.u.})]\times I(\mathrm{in\, W.m}^{-2})\,,
\end{equation}
and to the imaginary part of the polarizability corresponds the photon-scattering rate
\begin{equation}
\Gamma(\mathrm{in\, s}^{-1})=\frac{4\pi a_{0}^{3}}{\hbar c}\Im[\alpha(\mathrm{in\, a.u.})]\times I(\mathrm{in\, W.m}^{-2}).
\end{equation}
In what follows we assume the typical intensity of 1 GW.m$^{-2}$ (obtained for a laser power of 15 W and a gaussian beam waist of 100 $\textrm{\textmu m}$), which gives the potential $\overline{U}(\textrm{in\,\textmu K.GW}^{-1}\textrm{.m}^2)=0.22494655\times\Re[\alpha(\mathrm{in\, a.u.})]$ and the rate $\overline{\Gamma}(\mathrm{in\, s}^{-1}\textrm{.GW}^{-1}\textrm{.m}^2)=5.8900155\cdot10^{4}\times\Im[\alpha(\mathrm{in\, a.u.})]$.

\begin{figure}
\begin{centering}
\includegraphics[width=8cm]{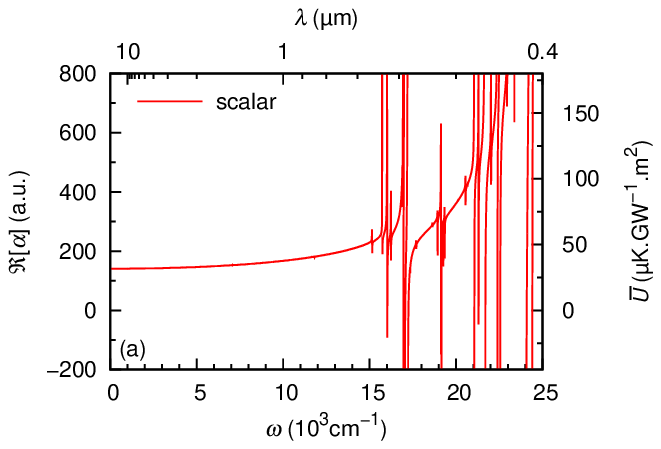}
\includegraphics[width=8cm]{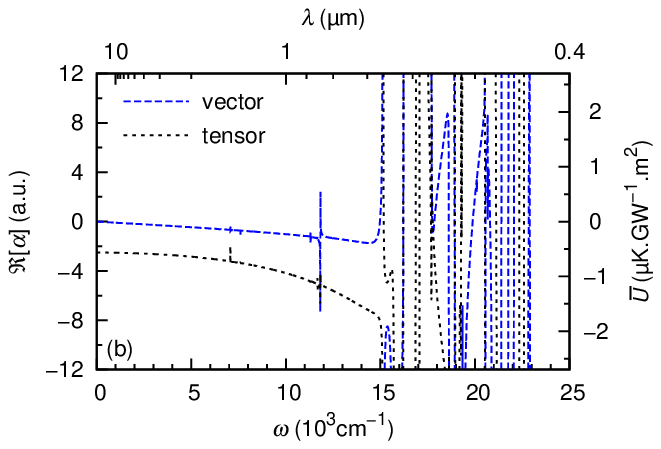}
\par\end{centering}
\caption{(Color online) Real part of the scalar (panel a), vector and tensor (panel b, resp.~dashed and dotted lines) polarizabilities in atomic units and corresponding trapping potentials obtained for an intensity of 1 GW.m$^{-2}$, as functions of the trapping frequency $\omega$ (or wavelength $\lambda$).
\label{fig:pola-real-freq}}
\end{figure}

\begin{table}
\caption{Real and imaginary parts of the scalar, vector and tensor polarizabilities in atomic units, at $\omega=0$ and 9398 cm$^{-1}$ ($\lambda=1064$ nm), compared with available literature values.
\label{tab:pola-re-im-freq}}
\begin{ruledtabular}
\begin{tabular}{ccc}
$\omega$ (cm$^{-1})$ & 0 & 9398\tabularnewline
\hline
$\Re[\alpha_{\mathrm{scal}}]$ & 141 & 164 \tabularnewline
 & 153 \cite{lide2012} & - \tabularnewline
 & 166 \cite{chu2007} & - \tabularnewline
$\Re[\alpha_{\mathrm{vect}}]$ & 0 & -0.943 \tabularnewline
$\Re[\alpha_{\mathrm{tens}}]$ & -2.52 & -3.93 \tabularnewline
 & -2.73 \cite{chu2007} & - \tabularnewline
\hline
$\Im[\alpha_{\mathrm{scal}}]$ & $1.51\cdot10^{-6}$ & $2.34\cdot10^{-6}$ \tabularnewline
$\Im[\alpha_{\mathrm{vect}}]$ & 0 & $-1.74\cdot10^{-6}$ \tabularnewline
$\Im[\alpha_{\mathrm{tens}}]$ & $-4.21\cdot10^{-7}$ & $-6.90\cdot10^{-7}$ \tabularnewline
\end{tabular}
\end{ruledtabular}
\end{table}

On Fig.~\ref{fig:pola-real-freq} we plot the real part of the erbium polarizabilities and the corresponding trapping potentials as functions of the laser frequency (in cm$^{-1}$) and wavelength (in nm). We see a dense pattern of resonances for $\omega\ge11000$ cm$^{-1}$. But most of them are narrow, which corresponds to weak transitions from the ground state, and the background profile of the polarizabilities is inherited from the strong lines. In Tab.~\ref{tab:pola-re-im-freq} we focus on two frequencies: $\omega=0$, to compare our results to the literature; and $\omega=9398$ cm$^{-1}$ ($\lambda=1064$ nm) a widespread laser-trapping frequency which is in the case of erbium far from any resonance. Our scalar and tensor polarizabilities are in good agreement with Refs.~\cite{lide2012, chu2007} which were calculated with different methods.

But the most striking feature is that the vector and tensor contributions are found extremely small compared to the scalar contribution. It means that the trapping potential exerted on erbium atoms is almost isotropic, in a sense that it does not depend on the respective orientation of the electronic cloud and the light polarization. One possible explanation to that phenomenon is the following: the anisotropic response to the trapping light should be due to the electrons of the unfilled $4f$ shell; but the latter is so contracted that the anisotropy is by far dominated by the isotropic response of the outermost $6s$ electrons.

\begin{figure}
\begin{centering}
\includegraphics[width=8cm]{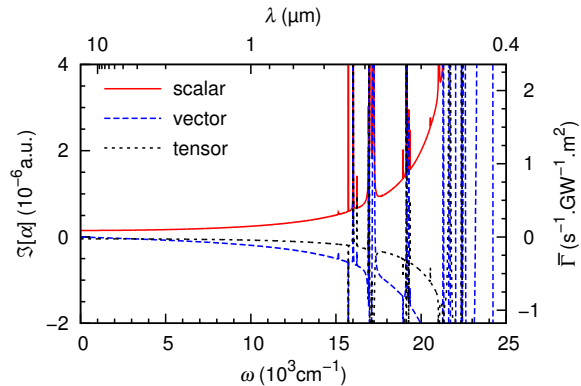}
\par\end{centering}
\caption{(Color online) Imaginary part of the scalar, vector and tensor polarizabilities (resp.~solid, dashed and dotted lines) in atomic units and corresponding photon-scattering rates obtained for an intensity of 1 GW.m$^{-2}$, as functions of the trapping frequency $\omega$ (or wavelength $\lambda$).
\label{fig:pola-imag-freq}}
\end{figure}

\begin{figure}
\begin{centering}
\includegraphics[width=8cm]{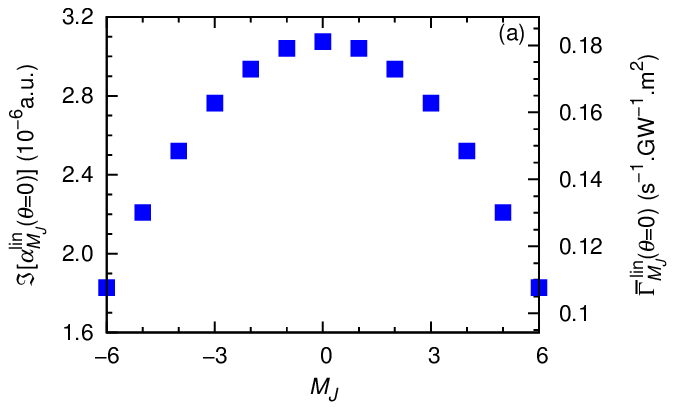}
\includegraphics[width=8cm]{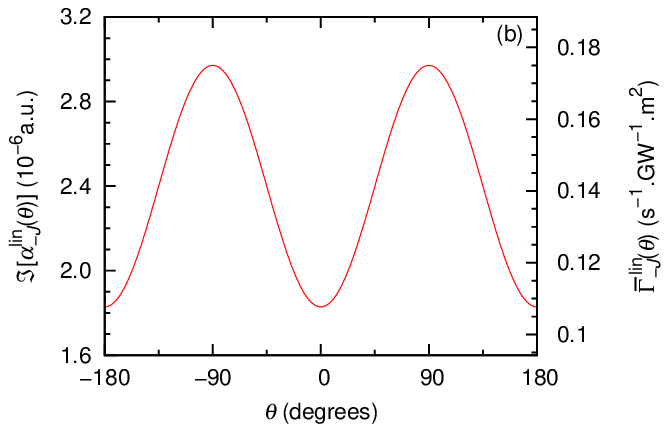}
\par\end{centering}
\caption{(Color online) Imaginary part of the polarizability in atomic units and corresponding photon-scattering rates obtained for a 1-GW.m$^{-2}$ linearly-polarized 1064-nm trapping light (see Eq.~(\ref{eq:Gamma})). Panel (a) corresponds to a polarization axis parallel to the quantization axis ($\theta=0^{\circ}$) and is as function of the Zeeman sublevel $M_{J}$; while panel (b) is for the lowest sublevel $M_{J}=-J$ and as a function of $\theta$.
\label{fig:pola-imag-theta-MJ}}
\end{figure}

The situation is drastically different for the imaginary part, for which the scalar, vector and tensor polarizabilities are of the same order of magnitude (see Fig.~\ref{fig:pola-imag-freq} and Tab.~\ref{tab:pola-re-im-freq}). The corresponding scaled photon-scattering rates are $\sim0.1$ s$^{-1}$.GW$^{-1}$.m$^{2}$. After a cycle of absorption and spontaneous emission, a fraction of the atoms are too hot to be kept in the trap. Therefore the atomic lifetime in the trap will strongly depend on the orientation between the electronic and the light polarization. This is illustrated on Figure \ref{fig:pola-imag-theta-MJ} where the reduced photon-scattering rate $\overline{\Gamma}_{-J}^{\mathrm{\,lin}}$ for the lowest Zeeman sublevel is plotted as a function of the angle $\theta$ between the linearly-polarized electric field and the quantization axis (given by an external magnetic field). Due to the negative sign of $\Im[\alpha_{\mathrm{tens}}]$ the rate is the smallest, and so the trap is the most stable, in the colinear configuration ($\theta=0$ or $180^{\circ}$). A similar behavior is observed as a function of $M_{J}$ for a fixed angle $\theta=0^{\circ}$.

\section{Concluding remarks \label{sec:conclu}}

In this article we give a new theoretical interpretation of the spectrum of neutral erbium, which enables us to characterize the optical trapping of ultracold erbium atoms. We find 9 unobserved levels which are accessible from the ground state through electric-dipole transition. We obtain a list of transition energies and transition dipole moments that we use to calculate the real and imaginary parts of the scalar, vector and tensor contributions to the ground-state polarizability. 
Although erbium is a non-spherically-symmetric atom, we show that the trapping potential exerted by an infrared laser is essentially isotropic, in the sense that it depends neither on the light polarization nor on the atomic Zeeman sublevel. In contrast, the photon-scattering rate exhibits an anisotropic behavior, since the vector and tensor contributions to the imaginary part of the polarizability are of the same order of magnitude as the scalar contribution. Calculations made with different transition energy and dipole moments including experimental ones show the same trends. Ongoing experiments in the Innsbruck group should allow us to check those results.

The anisotropy of the photon-scattering rate opens the possibility to control the heating or the losses in the trap with an appropriate light polarization. The dependence of the photon-scattering rate on the atomic sublevel also results in different trap lifetimes for different Feshbach-molecular states of Er$_2$ which are a current subject of interest.

Our calculations of polarizabilities are also relevant to characterize the long-range interactions between two erbium atoms. For non polarized atoms, \textit{i.e.}~not in a given Zeeman sublevel, the isotropic van der Waals coefficient $C_6^\mathrm{iso}$ can be calculated using the London formula $C_6^\mathrm{iso} = -(3/\pi) \int_0^{+\infty}d\omega\Re[\alpha_\mathrm{scal}(i\omega)]^2$, where $\alpha_\mathrm{scal}(i\omega)$ is the scalar polarizability at imaginary frequency, which gives for ground-state erbium $C_6^\mathrm{iso}=1760$ a.u.. 
In the case of polarized atoms, the $C_6$ coefficients also have an anisotropic contribution, which is very weak compared to the isotropic one, since it is proportional to $\int_0^{+\infty}d\omega \Re[\alpha_\mathrm{scal}(i\omega)] \Re[\alpha_\mathrm{tens}(i\omega)] = -16.6$ a.u.~and $\int_0^{+\infty}d\omega \Re[\alpha_\mathrm{tens}(i\omega)]^2 = 0.265$ a.u.. The $C_6$ coefficients between two polarized $^3H_6$ erbium atoms thus range from 1741 to 1766 a.u., and show a similar variation to dysprosium atoms \cite{kotochigova2011}.

The knowledge of the polarizabilities of excited states, and in particular of so-called {}``magic" frequencies or wavelengths, \textit{i.e.}~the wavelengths for which polarizabilities of the ground state and of a given excited state are equal, is of strong importance for precision measurements \cite{porsev2004, takamoto2005, ludlow2008}. In Ref.~\cite{dzuba2011}, the authors calculate magic wavelengths for dysprosium atoms in a non-polarized light. In the case of a polarized light, our preliminary calculations show that the anisotropy of trapping potential for erbium excited states tends to be larger than for the ground state. For example, at $\lambda=1064$ nm the polarizabilities of the level $J=7$ at 15847 cm$^{-1}$ are $\Re[\alpha_\mathrm{scal}]\approx 130$ a.u.~and $\Re[\alpha_\mathrm{scal}]\approx -60$ a.u.. So for the $M_J=-7$ sublevel in a linearly-polarized electric field, $\Re[\alpha_{M_J=-7}]$ varies from 70 a.u.~for $\theta=0^\circ$ to 160 a.u.~for $\theta=90^\circ$ [see Eq.~(\ref{eq:U-lin})].
This opens the possibility of a better control of the trapping conditions, by tuning both the laser wavelength and the polarization angle, as recently shown for diatomic molecules \cite{neyenhuis2012}.

\section*{Acknowledgments}
Enlightening discussions with the members of the experimental team of Francesca Ferlaino in Innsbruck, in particular Kiyotaka Aikawa, Simon Baier, Albert Frisch and Michael Mark, are gratefully acknowledged. ML is grateful to Eliane Luc for her guidance in theory of atomic structure. JFW acknowledges Laboratoire Aim\'e Cotton for its hospitality. Laboratoire Aim\'e Cotton is a member of \textit{Institut Francilien de Recherche sur les Atomes Froids} (www.ifraf.org).


\appendix
\section{Details of the atomic-structure calculations}

This appendix gives details on the calculation of the erbium atomic spectrum. Tables \ref{tab:parev} and \ref{tab:parev-ci} contains the optimal set of atomic parameters used for the last call of the diagonalization program RCG. For a given atomic parameter $P$, we multiply the HFR value by a scaling factor $SF(P)$, to obtain the input for the first call of RCG. The scaling factors $SF(P)$ given in Table \ref{tab:parev} and \ref{tab:parev-ci} were taken from our previous work on Er II \cite{wyart2009}. In addition to $E_\mathrm{av}$, $F^k(n\ell nè\ell')$, $G^k(n\ell nè\ell')$, $\zeta_{n\ell}$ and $R^k(n\ell n'\ell',n''\ell''n'''\ell''')$, the presence of effective parameters for accounting CI second order effects of far configurations. As explained in \cite{cowan1981, judd1985}, those parameters are $\alpha$, $\beta$ and $\gamma$ for the configurations $4f^{11}$ and $4f^{12}$ and Slater-forbidden parameters $F^{1}(4f,5d)$, $G^2(4f,5d)$ and $G^4(4f,5d)$ for the configurations with open $4f$ and $5d$ subshells. Due to the lack of HFR evaluations, initial values for effective parameters are derived from semi-empirical comparisons with similar spectra. Finally Tables \ref{tab:niver1e} and \ref{tab:niver1o} contain the characteristics of the calculated even- and odd-parity levels of Er I respectively.

\begin{longtable}{lrrrrrrrrr}

\caption{\label{tab:parev} Fitted one-configuration parameters 
(in cm$^{-1}$) for odd-parity configurations of Er I compared 
with HFR radial integrals. The scaling factors are 
$SF(P)=P_\mathrm{fit}/P_\mathrm{HFR}$, except for $E_\mathrm{av}$ 
when they are $P_{fit}-P_{HFR}$. 
The HFR values of $E_\mathrm{av}$ parameters are relative to the ground state
configuration $4f^{12}6s^2$ taken as zero value.
Some parameters are constrained 
to vary in a constant ratio $r_n$, indicated in the second column 
except if {}'fix' appears in the second or in the uncertainty columns. In this case, 
the parameter $P$ is not adjusted.}
\\
\hline \hline
 & &   \multicolumn{4}{c}{4f$^{11}5d6s^2$} & \multicolumn {4}{c}{4f$^{11}5d^26s$} \\
\cline{3-6} \cline{7-10}
\endfirsthead
\caption{Fitted parameters in Er I (continued)} \\
\hline
 & & \multicolumn{4}{c}{Fitted parameters} & \multicolumn{4}{c}{Fitted parameters} \\
\cline{3-6} \cline{7-10}
 Param. $P$      &  Cons.   & $P_\mathrm{fit}$ & Unc. & $P_\mathrm{HFR}$   & $SF$ &  $P_\mathrm{fit}$  & Unc.   & $P_\mathrm{HFR}$   & $SF$ \\
\hline
\endhead                                                                  
 Param. $P$      &  Cons.   & $P_\mathrm{fit}$ & Unc. & $P_\mathrm{HFR}$ &  $SF$  &  $P_\mathrm{fit}$ & Unc.  & $P_\mathrm{HFR}$ &  $SF$ \\
\hline
 $E_\mathrm{av}$ &          &           46389  &  68  &            6742  & 39647  &             65582 &   74  &            23334 & 42248 \\
 $F^2(4f4f)$     &  $r_{1}$ &           97984  & 387  &          128939  & 0.760  &             97812 &  387  &           128712 & 0.760 \\
 $F^4(4f4f)$     &  $r_{2}$ &           69490  & 308  &           80847  & 0.860  &             69360 &  307  &            80696 & 0.860 \\
 $F^6(4f4f)$     &  $r_{3}$ &           49446  & 631  &           58150  & 0.850  &             49351 &  630  &            58039 & 0.850 \\
 $\alpha$        &  $r_{4}$ &            21.0  &   2  &                  &        &              21.0 &    2  &                  &       \\
 $\beta$         &      fix &            -650  &      &                  &        &              -650 &       &                  &       \\
 $\gamma$        &      fix &            2000  &      &                  &        &              2000 &       &                  &       \\
 $F^2(5d5d)$     &          &                  &      &                  &        &             21541 &  323  &            32674 & 0.659 \\
 $F^4(5d5d)$     &          &                  &      &                  &        &             16590 &  611  &            20683 & 0.802 \\
 $\zeta_{4f}$    &  $r_{5}$ &            2381  &   4  &           2428   & 0.981  &              2379 &    4  &             2426 & 0.981 \\ 
 $\zeta_{5d}$    &  $r_{6}$ &             803  &   9  &            948   & 0.847  &               665 &    7  &              785 & 0.847 \\  
 $F^1(4f5d)$     &  $r_{7}$ &             671  &  70  &                  &        &               671 &   70  &                  &       \\ 
 $F^2(4f5d)$     &  $r_{8}$ &           15594  & 246  &           20265  & 0.770  &             13496 &  213  &            17539 & 0.769 \\ 
 $F^4(4f5d)$     &  $r_{9}$ &           10737  & 353  &            9189  & 1.168  &              9123 &  300  &             7807 & 1.169 \\
 $G^1(4f5d)$     & $r_{10}$ &            4997  & 121  &            8711  & 0.574  &              4277 &  104  &             7456 & 0.574 \\        
 $G^2(4f5d)$     & $r_{11}$ &            1238  & 291  &                  &        &              1238 &  291  &                  &       \\        
 $G^3(4f5d)$     & $r_{12}$ &            6103  & 286  &            6893  & 0.885  &              5170 &  242  &             5840 & 0.885 \\
 $G^4(4f5d)$     & $r_{13}$ &            1353  & 470  &                  &        &              1353 &  470  &                  &       \\        
 $G^5(4f5d)$     & $r_{14}$ &            4034  & 278  &            5204  & 0.775  &              3406 &  235  &             4394 & 0.775 \\
 $G^3(4f6s)$     & $r_{15}$ &                  &      &                  &        &              1080 &   90  &             1486 & 0.727 \\ 
 $G^2(5d6s)$     & $r_{17}$ &                  &      &                  &        &             12159 &  246  &            19202 & 0.633 \\
  \\                                                                                          
\hline \hline
 & & \multicolumn {4}{c}{4f$^{12}6s6p$} & \multicolumn {4}{c}{4f$^{12}5d6p$} \\
\cline{3-6} \cline{7-10}
     Param. $P$  &    Cons. & $P_\mathrm{fit}$ & Unc.  & $P_\mathrm{HFR}$ &  $SF$ & $P_\mathrm{fit}$ &  Unc.  & $P_\mathrm{HFR}$ &  $SF$ \\
\hline                                                                                             
 $E_\mathrm{av}$ &          &           36314  &   24  &            15491 & 20823 &            61570 &   fix  &            38570 & 23000 \\
 $F^2(4f4f)$     &  $r_{1}$ &           92534  &  366  &           121767 & 0.760 &            92314 &   365  &           121473 & 0.760 \\
 $F^4(4f4f)$     &  $r_{2}$ &           65336  &  290  &            76015 & 0.860 &            65168 &   289  &            75822 & 0.860 \\
 $F^6(4f4f)$     &  $r_{3}$ &           46412  &  592  &            54582 & 0.850 &            46294 &   591  &            54441 & 0.850 \\
 $\alpha$        &  $r_{4}$ &            21.0  &    2  &                  &       &             21.0 &     2  &                  &       \\
 $\beta$         &    fix   &            -650  &       &                  &       &             -650 &        &                  &       \\
 $\gamma$        &    fix   &            2000  &       &                  &       &             2000 &        &                  &       \\
 $\zeta_{4f}$    &  $r_{5}$ &            2242  &    4  &             2286 & 0.981 &             2242 &     4  &             2284 & 0.981 \\ 
 $\zeta_{5d}$    &  $r_{6}$ &                  &       &                  &       &              463 &     5  &              547 & 0.846 \\ 
 $\zeta_{6p}$    & $r_{18}$ &            1496  &   18  &             1035 & 1.445 &             1107 &    13  &              766 & 1.445 \\  
 $F^1(4f5d)$     &  $r_{7}$ &                  &       &                  &       &              671 &    70  &                  &       \\           
 $F^2(4f5d)$     &  $r_{8}$ &                  &       &                  &       &            11010 &   174  &            14308 & 0.703 \\  
 $F^4(4f5d)$     &  $r_{9}$ &                  &       &                  &       &             7345 &   241  &             6286 & 1.168 \\ 
 $F^1(4f6p)$     &    fix   &             100  &       &                  &       &              150 &        &                  &       \\  
 $F^2(4f6p)$     & $r_{19}$ &            3698  &  155  &             3267 &  1.13 &             2943 &   123  &             2610 & 1.13  \\
 $F^2(6p5d)$     &      fix &                  &       &                  &       &            11470 &        &            14438 & 0.80  \\           
 $G^1(4f5d)$     &      fix &                  &       &                  &       &             3898 &        &             6652 & 0.724 \\        
 $G^2(4f5d)$     &      fix &                  &       &                  &       &             1092 &        &                  &       \\        
 $G^3(4f5d)$     &      fix &                  &       &             9357 & 0.898 &             4397 &        &             4913 & 0.895 \\
 $G^4(4f5d)$     &      fix &                  &       &                  &       &             1028 &        &                  &       \\      
 $G^5(4f5d)$     &      fix &                  &       &                  &       &             2761 &        &             3625 & 0.762 \\
 $G^3(4f6s)$     & $r_{15}$ &            1210  &  101  &             1665 & 0.727 &                  &        &                  &       \\  
 $G^2(4f6p)$     & $r_{19}$ &             843  &   35  &              748 &  1.13 &              643 &        &              568 & 1.13  \\        
 $G^4(4f6p)$     & $r_{19}$ &             733  &   31  &              650 &  1.13 &              556 &        &              491 & 1.13  \\
 $G^1(6s6p)$     &          &           12843  &   66  &            23373 & 0.549 &                  &        &                  &       \\        
 $G^1(6p5d)$     &      fix &                  &       &                  &       &             7880 &        &            13133 & 0.60  \\        
 $G^3(6p5d)$     &      fix &                  &       &                  &       &             5052 &        &             8420 & 0.60  \\
\hline \hline
\end{longtable}

\begin{longtable}{lrrrrr}
\caption{\label{tab:parev-ci} Same as Table~\ref{tab:parev} for 
configuration-interaction parameters.} \\
\hline \hline
\endfirsthead
\caption{Fitted parameters in Er I (continued)} \\
\hline \hline
 Param. $P$         &    Cons. & $P_\mathrm{fit}$ & Unc. & $P_\mathrm{HFR}$   & $SF$ \\
\hline
\endhead
 & & \multicolumn{4}{c}{$4f^{11}5d6s^2-4f^{11}5d^26s$} \\
\cline{3-6}
 Param. $P$         &    Cons. & $P_\mathrm{fit}$ & Unc. & $P_\mathrm{HFR}$   & $SF$ \\
\hline
 $R^{2}(4f6s,4f5d)$ &      fix &   -710 &      &   -938 & 0.757 \\
 $R^{3}(4f6s,4f5d)$ &      fix &   1002 &      &    770 & 0.757 \\
 $R^{2}(5d6s,5d5d)$ & $r_{17}$ & -13919 &  282 & -21982 & 0.633 \\
\\
 & & \multicolumn{4}{c}{$4f^{11}5d6s^2-4f^{12}6s6p$} \\
\cline{3-6}
 $R^{1}(5d6s,4f6p)$ & $r_{20}$ &  -3163 &   96 &  -6878 & 0.46  \\
 $R^{3}(5d6s,6p4f)$ & $r_{20}$ &   -678 &   21 &  -1474 & 0.46  \\
\\
 & & \multicolumn{4}{c}{$4f^{11}5d^26s-4f^{12}6s6p$} \\   
\cline{3-6}
 $R^{1}(5d5d,4f6p)$ &      fix &   2813 &      &   3715 & 0.757 \\
 $R^{3}(5d5d,4f6p)$ &      fix &    753 &      &    994 & 0.757 \\
\\
 & & \multicolumn{4}{c}{$4f^{11}5d^26s-4f^{12}5d6p$} \\   
\cline{3-6}
 $R^{1}(5d6s,4f6p)$ &      fix &  -2966 &      &  -5932 & 0.5   \\
 $R^{3}(5d6s,4f6p)$ &      fix &   -660 &      &  -1320 & 0.5   \\
\hline \hline
\end{longtable}

\begin{longtable}{lrrrrrr}
\caption{Comparison of energies $E$ and Land\'e factors $g_L$ 
of Er I even-parity 
levels. The superscript {}''exp" stands for experimental values
which are taken from \cite{martin1978}. The superscript {}''th" 
stands for the theoretical values from the present parametric 
calculations. Note that 
$\Delta E = E^\mathrm{exp} - E^\mathrm{th}$.}
\label{tab:niver1e}
\\
 \hline \hline
Configuration &  Term & $J$ & $E^\mathrm{exp}$ & $g_L^\mathrm{exp}$ & $\Delta E$ & $g_L^\mathrm{th}$ \\
 \hline                                                                   
\endfirsthead
\caption{Even parity levels of Er I (continued)}\\
 \hline
\endhead
$4f^{12}6s^2$ & $^3H$ &   6 &        0         &            1.16381 &         -9 &             1.166 \\
$4f^{12}6s^2$ & $^3F$ &   4 &     5035.193     &            1.147   &         49 &             1.141 \\ 
$4f^{12}6s^2$ & $^3H$ &   5 &     6958.329     &            1.031   &        -15 &             1.033 \\
$4f^{12}6s^2$ & $^3H$ &   4 &    10750.982     &            0.936   &       -116 &             0.945 \\
$4f^{12}6s^2$ & $^3F$ &   3 &    12377.534     &            1.065   &        -53 &             1.084 \\
$4f^{12}6s^2$ & $^3F$ &   2 &    13097.906     &            0.750   &         25 &             0.739 \\
 \hline \hline
\end{longtable}

\begin{longtable}{rrrrlllrrrr}
\caption{Same as \ref{tab:niver1e} for Er I odd-parity levels 
with electric dipole decay to the ground level. 
The theoretical values $E^\mathrm{th}$, the factors 
$g_L^\mathrm{th}$ and the percentage of calculated configurations 
are derived by means of the RCG code with the parameter set 
reported in Table \ref{tab:parev}.
In the configuration notations, $A$ stands for $4f^{12}$, 
$B$ for $4f^{11}$, $ds^2$ for $5d6s^2$, $sp$ for $6s6p$, 
$d^2s$ for $5d^26s$ and $dp$ for $5d6p$.
The lower-case letters or Arabic numbers appearing in the seventh column correspond to different intermediate coupling schemes \cite{cowan1981}.}
\label{tab:niver1o}                             
\endfirsthead
\caption{Odd parity levels of Er I (continued)} \\
\hline
 & & & & & Leading & \% leading & \multicolumn{4}{c}{Configuration}   \\
 $E^\mathrm{exp}$ \cite{martin1978} & $E^\mathrm{th}$ &  $\Delta E $ & $g_L^\mathrm{exp}$ & $g_L^\mathrm{th}$ & electr. & LS-coupling & \multicolumn{4}{c}{weight (\%)}  \\
 \cline{8-11}
   &   &   &   &  & config. & component & $B-ds^2$  & $B-d^2s$  & $A-sp$ & $A-dp$  \\
\hline
\endhead
\hline
 & & & & & Leading & \% leading & \multicolumn{4}{c}{Configuration}   \\
 $E^\mathrm{exp}$ \cite{martin1978} & $E^\mathrm{th}$ &  $\Delta E $ & $g_L^\mathrm{exp}$ & $g_L^\mathrm{th}$ & electr. & LS-coupling & \multicolumn{4}{c}{weight (\%)}  \\
 \cline{8-11}
   &   &   &   &  & config. & component & $B-ds^2$  & $B-d^2s$  & $A-sp$ & $A-dp$  \\
\hline
               &  $J=5$      &       &       &       &           &                           &       &       &       &         \\
    11401.197  &  11419.6    & -18   & 1.205 & 1.210 & $B-ds^2$  &   51 $B-ds^2 \, (^4I) ^5G$   &  97.5 &   2.3 &   0.3 &     0   \\ 
    15185.352  &  15258.6    &  -7   & 1.160 & 1.170 & $B-ds^2$  &   55 $B-ds^2 \, (^4I) ^3G$   &  96.7 &   2.3 &   1.0 &     0   \\ 
    17029.058  &  17023.5    &   6   & 1.150 & 1.136 & $B-ds^2$  &   37 $B-ds^2 \, (^4I) ^5H$   &  97.3 &   2.3 &   0.3 &     0   \\   
    17347.860  &  17314.6    &  33   & 1.175 & 1.177 & $A-sp$    &   33 $A-sp   \, (^3H) ^3Ga$  &   0.3 &   1.2 &  98.1 &   0.4   \\    
    19201.343  &  19250.1    & -49   & 1.060 & 1.059 & $A-sp$    &   25 $A-sp   \, (^3H) ^1H$   &  16.2 &   0.9 &  82.6 &   0.3   \\     
    19563.116  &  19383.6    &  180  & 0.990 & 0.990 & $B-ds^2$  &   26 $B-ds^2 \, (^4I) ^5I$   &  81.1 &   2.1 &  16.8 &   0.1   \\    
    20917.276  &  20790.1    &  127  & 0.980 & 0.980 & $B-ds^2$  &   23 $B-ds^2 \, (^4I) ^5K$   &  97.6 &   2.2 &   0.2 &     0   \\    
    21392.817  &  21419.2    &  -26  & 1.005 & 1.019 & $B-ds^2$  &   18 $B-ds^2 \, (^4I) ^3I$   &  95.6 &   2.6 &   1.7 &   0.1   \\    
    22124.268  &  22136.3    & -12   & 1.285 & 1.264 & $A-sp$    &   28 $A-sp   \, (^3F) ^5F$   &   0.3 &  15.6 &  83.7 &   0.4   \\   
    22450.111  &  22571.5    &  -121 & 1.360 & 1.370 & $B-d^2s$  &   35 $B-d^2s \, (^4I) ^7F$   &   1.4 &  82.7 &  15.8 &   0.1   \\ 
    22672.766  &  22651.3    &   21  & 1.040 & 1.040 & $B-ds^2$  &   22 $B-ds^2 \, (^4I) ^5K$   &  92.0 &   3.7 &   4.0 &   0.2   \\  
    23447.079  &  23475.0    &  -28  & 1.080 & 1.084 & $A-sp$    &   23 $A-sp   \, (^3H) ^5I$   &   1.0 &   1.9 &  96.8 &   0.4   \\  
    23855.654  &  23878.9    &  -23  & 1.140 & 1.178 & $A-sp$    &   26 $A-sp   \, (^3F) ^5F$   &   1.7 &   1.9 &  96.0 &   0.4   \\ 
    23885.406  &  23903.7    & -18   & 1.100 & 1.058 & $A-sp$    &   22 $A-sp   \, (^3H) ^5I$   &   4.6 &   1.7 &  93.3 &   0.5   \\  
    24083.166  &  24055.6    &  28   & 1.128 & 1.132 & $A-sp$    &   46 $A-sp   \, (^3H) ^3Gb$  &  37.1 &   1.7 &  58.5 &   2.6   \\  
    25162.553  &  25170.9    &  -8   & 1.010 & 1.016 & $B-ds^2$  &   24 $B-ds^2 \, (^4I) ^3I$   &  74.9 &   3.5 &  20.5 &   1.1   \\     
    25364.012  &  25382.3    & -18   & 1.180 & 1.183 & $B-d^2s$  &   13 $B-d^2s \, (^4I) ^7Ha$  &   1.6 &  96.8 &   1.4 &   3.2   \\  
    25681.933  &  25598.0    &   84  & 1.175 & 1.142 & $B-ds^2$  &   20 $B-ds^2 \, (^4F) ^5F$   &  81.1 &   3.0 &  15.0 &   0.9   \\   
    26198.837  &  26145.5    &  53   & 1.045 & 1.069 & $A-sp$    &   48 $A-sp   \, (^3H) ^5H$   &   0.1 &   1.4 &  98.1 &   0.3   \\   
               &  27651.7    &       &       & 1.315 & $B-d^2s$  &   41 $B-d^2s \, (^4I) ^5Fb$  &   2.2 &  97.0 &   0.8 &   0.1   \\  
    27856.436  &  27825.9    &  31   & 1.095 & 1.145 & $B-d^2s$  &   15 $B-d^2s \, (^4I) ^7G$   &   8.0 &  87.7 &   4.1 &   0.2   \\    
    28026.045  &  28090.7    & -65   & 1.120 & 1.056 & $A-sp$    &   18 $A-sp   \, (^3H) ^5I$   &  12.7 &   9.4 &  77.5 &   0.4   \\      
    28129.803  &  28141.9    & -12   & 1.040 & 1.125 & $B-ds^2$  &   10 $B-ds^2 \, (^4G) ^5G$   &  67.6 &   8.6 &  23.6 &   0.3   \\    
    29272.207  &  29237.0    &  35   & 1.115 & 1.123 & $B-ds^2$  &   37 $B-ds^2 \, (^4F) ^5H$   &  95.1 &   2.7 &   2.1 &   0.1   \\     
    29550.807  &  29770.5    & -220  & 1.150 & 1.168 & $B-ds^2$  &   32 $B-ds^2 \, (^4F) ^5G$   &  69.7 &   4.6 &  25.0 &   0.7   \\    
    29794.862  &  29821.6    &  -27  & 1.100 & 1.131 & $A-sp$    &   14 $A-sp   \, (^3F) ^5F$   &  17.1 &   4.5 &  78.1 &   0.3   \\     
    29894.203  &  30064.0    & -170  & 1.195 & 1.126 & $B-d^2s$  &   17 $B-d^2s \, (^4I) ^7Ia$  &   1.8 &  90.5 &   7.3 &   0.4   \\    
    30380.282  &  30326.8    & 53    & 1.116 &       & $A-sp$    &   20 $A-sp   \, (^3F) ^3Gb$  &  15.6 &  27.3 &  53.9 &   3.3   \\ 
    30600.160  &  30768.9    & -169  & 1.195 & 1.093 & $B-d^2s$  &\,\,\,8 $B-d^2s \, (^4I) ^5Hc$  &   4.9 &  78.4 &  15.2 &   1.4   \\    
    31105.090  &  30988.2    &  117  & 1.200 & 1.250 & $B-d^2s$  &   37 $B-d^2s \, (^4I) ^5Fa$  &   0.6 &  96.4 &   2.1 &   0.9   \\    
    31194.235  &  31185.2    &   9   & 1.135 & 1.128 & $B-ds^2$  &   16 $B-ds^2 \, (^4F) ^5H$   &  86.5 &   9.5 &   3.8 &   0.2   \\    
    31364.719  &  31360.7    &  4    & 1.235 & 1.232 & $A-sp$    &\,\,\,4 $A-sp   \, (^3F) ^5G$   &   2.5 &   5.1 &  92.1 &   0.4   \\     
    31442.927  &  31475.4    & -33   & 1.195 & 1.132 & $B-d^2s$  &   12 $B-d^2s \, (^4I) ^5Fa$  &   3.7 &  90.5 &   5.2 &   0.6   \\    
\hline      
  &   $J=6$ & & & & & & & &      \\ 
     7176.503  &    7185.5  &   -9  & 1.302 & 1.304 & $B-ds^2$  &    77 $B-ds^2 \, (^4I) ^5G$   & 98.2  &  1.8  &    0  &    0    \\      
    11799.778  &   11788.5  &   11  & 1.190 & 1.195 & $B-ds^2$  &    39 $B-ds^2 \, (^4I) ^5H$   & 96.9  &  2.5  &  0.5  &    0    \\      
    16070.095  &   16125.4  &  -55  & 1.200 & 1.169 & $B-ds^2$  &    42 $B-ds^2 \, (^4I) ^3H$   & 76.5  &  2.3  & 19.1  &  0.2    \\      
    16321.110  &   16347.4  &  -26  & 1.220 & 1.254 & $A-sp$    &    59 $A-sp   \, (^3H) ^5G$   & 14.6  &  1.4  & 83.7  &  0.3    \\      
    17073.800  &   17063.7  &   10  & 1.070 & 1.069 & $A-sp$    &    27 $A-sp   \, (^3H) ^3Ia$  &  2.9  &  1.1  & 95.6  &  0.4    \\     
    17456.383  &   17461.9  &   -6  & 1.070 & 1.058 & $B-ds^2$  &    23 $B-ds^2 \, (^4I) ^5I$   & 96.5  &  2.7  &  0.8  &    0    \\      
    19326.598  &   19273.3  &   53  & 1.180 & 1.175 & $A-sp$    &    31 $A-sp   \, (^3H) ^5H$   &  0.5  &  0.8  & 98.3  &  0.3    \\      
    19508.432  &   19461.5  &   47  & 0.960 & 0.960 & $B-ds^2$  &    35 $B-ds^2 \, (^4I) ^5K$   & 96.7  &  2.5  &  0.8  &    0    \\     
    20166.130  &   20213.2  &  -47  & 1.485 & 1.475 & $B-d^2s$  &    78 $B-d^2s \, (^4I) ^7F$   &  0.1  & 99.9  &    0  &    0    \\      
    20737.723  &   20659.6  &   78  & 0.855 & 0.853 & $B-ds^2$  &    49 $B-ds^2 \, (^4I) ^5L$   & 97.6  &  2.3  &  0.2  &    0    \\     
    21701.885  &   21786.2  &  -84  & 1.055 & 1.045 & $B-ds^2$  &    25 $B-ds^2 \, (^4I) ^3I$   & 91.3  &  2.9  &  5.5  &  0.2    \\      
    22583.504  &   22501.1  &   82  & 1.130 & 1.137 & $B-ds^2$  &    36 $B-ds^2 \, (^4F) ^5G$   & 95.6  &  2.5  &  1.8  &  0.1    \\      
    23311.577  &   23311.6  &    0  & 1.250 & 1.267 & $B-d^2s$  &    22 $B-d^2s \, (^4I) ^7Ha$  &  0.4  & 97.3  &  2.3  &  0.1    \\    
    23831.359  &   23820.4  &   11  & 1.250 & 1.248 & $A-sp$    &    56 $A-sp   \, (^3F) ^5G$   &  0.3  &  2.9  & 96.4  &  0.4    \\    
    24246.146  &   24215.8  &   30  & 1.085 & 1.098 & $A-sp$    &    43 $A-sp   \, (^3H) ^5I$   &  1.3  &  1.6  & 96.7  &  0.4    \\   
    24457.139  &   24492.3  &  -35  & 1.050 & 1.054 & $B-ds^2$  &    24 $B-ds^2 \, (^4F) ^3H$   & 84.4  &  2.7  & 12.3  &  0.6     \\    
    25268.259  &   25308.9  &  -40  & 1.185 & 1.166 & $B-d^2s$  &    17 $B-d^2s \, (^4I) ^7G$   &  6.3  & 91.8  &  1.7  &  0.1    \\    
    25392.779  &   25419.3  &  -27  & 1.075 & 1.072 & $B-ds^2$  &    21 $B-ds^2 \, (^4F) ^5G$   & 82.9  &  9.3  &  7.4  &  0.4    \\    
    25880.274  &   26070.5  & -190  & 1.150 & 1.156 & $A-sp$    &    41 $A-sp   \, (^3H) ^3Hb$  & 19.9  &  2.3  & 75.7  &  2.1    \\  
    26237.004  &   26178.0  &   59  & 1.160 & 1.158 & $A-sp$    &    36 $A-sp   \, (^3H) ^3Ha$  & 15.3  &  2.7  & 80.1  &  1.9    \\ 
    27582.017  &   27490.5  &   91  & 1.120 & 1.113 & $B-d^2s$  &    12 $B-d^2s \, (^4I) ^5Ha $ &  0.5  & 98.9  &  0.4  &  0.2    \\ 
    27879.416  &   27996.0  & -117  & 1.175 & 1.147 & $B-ds^2$  &    23 $B-ds^2 \, (^4G) ^5G$   & 90.0  &  7.9  &  2.0  &  0.1    \\  
    28854.941  &   28902.8  &  -48  & 1.190 & 1.208 & $B-d^2s$  &    22 $B-d^2s \, (^4I) ^5Gb$  &  3.1  & 96.0  &  0.8  &  0.1    \\ 
    29152.796  &   29118.8  &   34  & 1.175 & 1.192 & $B-d^2s$  &    15 $B-d^2s \, (^4I) ^7Hb$  &  0.6  & 99.0  &  0.1  &  0.2    \\  
    29561.425  &   29584.5  &  -24  & 1.130 & 1.126 & $A-sp$    &    25 $A-sp   \, (^3F) ^5G$   &  0.2  &  3.1  & 96.2  &  0.4    \\  
               &   29718.1  &       &       & 1.114 & $B-d^2s$  &\,\,\,9 $B-d^2s\, (^4I) ^3Ic$  & 16.1  & 81.3  &  2.3  &  0.3    \\  
    30007.369  &   30051.0  &  -44  & 1.090 & 1.092 & $B-ds^2$  &    14 $B-ds^2 \, (^2H) ^1I2$  & 67.8  & 31.2  &  0.9  &  0.1    \\  
    30088.200  &   30169.1  &  -81  & 1.120 & 1.126 & $B-d^2s$  &    17 $B-d^2s \, (^4I) ^7Ka$  & 12.1  & 87.3  &  0.4  &  0.2    \\ 
               &   30702.6  &       &       & 1.268 & $B-d^2s$  &    26 $B-d^2s \, (^4I) ^5Ga$  &  1.6  & 97.6  &  0.3  &  0.4    \\  
    30765.720  &   30771.7  &   -6  & 1.205 & 1.205 & $B-ds^2$  &    51 $B-ds^2 \, (^4F) ^5H$   & 95.6  &  4.4  &  0.1  &    0    \\   
    31205.223  &   31264.5  &  -59  & 1.100 & 1.090 & $B-d^2s$  &    14 $B-d^2s \, (^4I) ^7Ia$  &  0.6  & 99.1  &  0.1  &  0.2    \\   
    31823.748  &   31706.0  &  118  & 1.045 & 1.078 & $A-sp$    &    39 $A-sp   \, (^3H) ^3Ib$  &  6.7  & 27.7  & 61.9  &  3.7    \\
    31926.003  &   31939.5  &  -13  & 1.215 & 1.215 & $B-d^2s$  &    18 $B-d^2s \, (^4I) ^5Gd$  &  4.1  & 76.4  & 18.0  &  1.5    \\
\hline
  & $J=7$ &  & & & & & & & &  \\
     7696.956 &       7713.9 &  -17   & 1.266 & 1.262  & $B-ds^2$   &    78 $B-ds^2 \, (^4I) ^5H$    & 98.0  &  2.0  &    0  &    0    \\   
    11887.503 &      11937.5 &  -50   & 1.153 & 1.150  & $B-ds^2$   &    47 $B-ds^2 \, (^4I) ^3I$    & 96.6  &  2.9  &  0.5  &    0    \\   
    15846.549 &      15844.1 &    2   & 1.070 & 1.066  & $B-ds^2$   &    43 $B-ds^2 \, (^4I) ^5K$    & 96.6  &  2.4  &  0.9  &    0    \\  
    17157.307 &      17129.3 &   28   & 1.195 & 1.192  & $A-sp$     &    39 $A-sp   \, (^3H) ^5I$    &  1.9  &  1.1  & 96.6  &  0.4    \\    
    17796.139 &      17809.4 &  -13   & 1.110 & 1.107  & $B-ds^2$   &    48 $B-ds^2 \, (^4I) ^5I$    & 95.4  &  2.8  &  1.7  &  0.0    \\   
    18774.123 &      18737.9 &   36   & 0.965 & 0.967  & $B-ds^2$   &    45 $B-ds^2 \, (^4I) ^5L$    & 97.6  &  2.4  &    0  &    0    \\ 
    19125.253 &      19052.2 &   73   & 1.235 & 1.244  & $A-sp$     &    68 $A-sp   \, (^3H) ^5H$    &    0  &  1.0  & 98.6  &  0.4    \\  
    21168.430 &      21162.2 &    6   & 1.065 & 1.062  & $B-ds^2$   &    33 $B-ds^2 \, (^4I) ^3K$    & 95.9  &  2.9  &  1.1  &  0.1    \\  
    21787.932 &      21749.2 &   39   & 1.350 & 1.360  & $B-d^2s$   &    49 $B-d^2s \, (^4I) ^7G$    &  0.2  & 99.8  &    0  &    0    \\  
    23080.952 &      23046.6 &   34   & 1.010 & 1.011  & $B-ds^2$   &    40 $B-ds^2 \, (^4I) ^3L$    & 97.0  &  2.8  &  0.2  &    0    \\ 
    23364.853 &      23396.3 &  -31   & 1.225 & 1.226  & $B-d^2s$   &    24 $B-d^2s \, (^4I) ^7G$    &  0.3  & 99.6  &  0.2  &    0    \\ 
    24943.272 &      24946.1 &   -3   & 1.160 & 1.145  & $A-sp$     &    57 $A-sp   \, (^3H) ^3Ib$   &  3.9  &  6.5  & 84.6  &  4.9    \\ 
    25159.143 &      25167.9 &   -9   & 1.170 & 1.170  & $B-d^2s$   &    12 $B-d^2s \, (^4I) ^7Ha$   &  0.9  & 93.8  &  4.9  &  0.4    \\ 
    25598.286 &      25570.2 &   28   & 1.155 & 1.166  & $A-sp$     &    48 $A-sp   \, (^3H) ^5I$    &  0.4  &  2.0  & 97.1  &  0.5    \\ 
              &      25659.2 &        &       & 1.146  & $B-ds^2$   &    48 $B-ds^2 \, (^4F) ^5H$    & 96.6  &  3.1  &  0.3  &    0    \\ 
    27230.646 &      27134.8 &   96   & 1.135 & 1.113  & $B-d^2s$   &    12 $B-d^2s \, (^4I) ^1Ka$   &  0.2  & 99.5  &  0.1  &  0.2    \\ 
    27306.747 &      27432.5 & -126   & 1.225 & 1.243  & $B-d^2s$   &    25 $B-d^2s \, (^4I) ^7Hb$   &  0.1  & 99.8  &    0  &  0.1    \\ 
    28017.584 &      28087.3 &  -70   & 1.080 & 1.068  & $B-ds^2$   &    24 $B-ds^2 \, (^2H) ^3K2$   & 93.5  &  6.5  &    0  &    0    \\ 
              &      28306.8 &        &       & 1.220  & $B-d^2s$   &    19 $B-d^2s \, (^4I) ^7Hb$   &  1.8  & 98.0  &  0.1  &  0.1    \\ 
              &      29088.0 &        &       & 1.171  & $B-d^2s$   &    18 $B-d^2s \, (^4I) ^5Hb$   &  2.1  & 97.9  &    0  &    0    \\ 
              &      29781.0 &        &       & 1.069  & $B-d^2s$   &    15 $B-d^2s \, (^4I) ^1Kb$   &  1.8  & 98.0  &    0  &  0.2    \\   
              &      30127.6 &        &       & 1.215  & $B-d^2s$   &    21 $B-d^2s \, (^4I) ^5Hf$   & 12.0  & 87.9  &  0.1  &  0.1    \\  
              &      30353.5 &        &       & 1.117  & $B-d^2s$   &    17 $B-ds^2 \, (^2H) ^3I2$   & 45.1  & 54.6  &  0.2  &  0.1    \\ 
\hline
\end{longtable}

\section*{References}


\end{document}